\documentclass[journal]{IEEEtran}

\usepackage{amssymb}
\usepackage{amsmath,amsthm}
\usepackage{dsfont,amsfonts}
\usepackage{algorithm,algorithmic}
\usepackage{graphicx}  
\usepackage{epstopdf}
\usepackage{hyperref}
\usepackage{xcolor}
\usepackage{cite}
\usepackage{multirow}
\usepackage{booktabs}
\usepackage{multicol}
\usepackage{subfig}
\allowdisplaybreaks[4]
\usepackage{enumitem}
\setitemize[1]{itemsep=4.5pt,partopsep=1pt,parsep=\parskip,topsep=4.5pt}
\usepackage{setspace}
\newtheorem{theorem}{Theorem}
\newtheorem{remark}{Remark}

\newtheorem{definition}{Definition}

\def\subparagraph{} 
\usepackage{titlesec}
\titlespacing*{\section}{0pt}{*0.8}{*0.8}
\titlespacing{\subsection}{0pt}{*0.8}{*0.8}
\renewcommand{\thesubsubsection}{\arabic{subsubsection}}
\titleformat{\subsubsection}[runin]{\itshape}{\thesubsubsection)}{1em}{}
\titlespacing*{\subsubsection}{\parindent}{0pt}{*0.8}

\makeatletter
\def\thmhead@plain#1#2#3{%
  \thmname{#1}\thmnumber{\@ifnotempty{#1}{ }\@upn{#2}}%
  \thmnote{ {\the\thm@notefont#3}}}
\def\thm@space@setup{\thm@preskip=2pt
\thm@postskip=2pt}
\let\thmhead\thmhead@plain
\makeatother

\usepackage[acronym,nowarn,hyperfirst=false]{glossaries}
\setkeys{glslink}{hyper=false}
\newacronym{nlp}{\textsc{nlp}}{nonlinear programming}
\newacronym{itd}{ITD}{integrated transmission-distribution}
\newacronym{qp}{\textsc{qp}}{quadratic program}
\newacronym{ocd}{\textsc{ocd}}{Optimality Condition Decomposition}
\newacronym{app}{\textsc{app}}{Auxiliary Problem Principle}
\newacronym{sqp}{\textsc{sqp}}{Sequential Quadratic Programming}
\newacronym{sdp}{SDP}{semi-definite programming}
\newacronym{soc}{SOC}{second-order cone}
\newacronym{qcqp}{\textsc{qcqp}}{Quadratically Constrained Quadratic Program} 
\newacronym{rapidpf}{rapid\textsc{pf}}{rapid prototyping for distributed Power Flow}
\newacronym{admm}{\textsc{admm}}{Alternating Direction Method of Multipliers}
\newacronym{dqa}{\textsc{dqa}}{Diagonal Quadratic Approximation}
\newacronym{aladin}{\textsc{aladin}}{Augmented Lagrangian based Alternating Direction Inexact Newton method}
\newacronym{aladincor}{\textsc{aladin}-\textsc{cor}}{\acrshort{aladin} with second-order correction}
\newacronym{aladinstd}{\textsc{aladin}-\textsc{std}}{standard \acrshort{aladin}}
\newacronym{hgd}{\textsc{hgd}}{heterogeneous decomposition}
\newacronym{dcc}{\textsc{dcc}}{Distribution-Cost-Correction}
\newacronym{rmse}{\textsc{rmse}}{Root-Mean-Square Error}
\newacronym{pcc}{PPC}{point of common coupling}
\newacronym{vsc}{VSC}{voltage source converter}
\newacronym{mtdc}{MTDC}{multiterminal high voltage direct current}
\newacronym{hvdc}{HVDC}{high voltage direct current }
\newacronym{hvac}{HVAC}{high voltage alternating current }
\newacronym{lcc}{LCC}{line commutated converters}
\newacronym{lcc-hvdc}{LCC-HVDC}{\acrfull{lcc} based multiterminal high voltage direct current}
\newacronym{vsc-mtdc}{VSC-MTDC}{\acrfull{vsc} based multiterminal high voltage direct current}
\newacronym{igbt}{IGBT}{insulated gate bipolar transistor}
\newacronym{vsc-hvdc}{VSC-HVDC}{\acrfull{vsc} based high voltage direct current}
\newacronym{opf}{OPF}{optimal power flow}
\newacronym{bim}{BIM}{bus injection model}
\newacronym{bfm}{BFM}{branch flow model}
\newacronym{ders}{DERs}{distributed energy resources}

\newacronym{pu}{p.u.}{per-unit system}
\newacronym{pwm}{PWM}{pulse-width modulation}
\newacronym{licq}{LICQ}{linear independence constraint qualification} 
\newacronym{sosc}{SOSC}{second order sufficient condition} 
\newacronym{scc}{SCC}{strict complementarity conditions} 
\newacronym{kkt}{KKT}{Karush–Kuhn–Tucker} 

\newcommand{\norm}[1]{\left\lVert#1\right\rVert}

\newcommand{\pqp}{p^\textsc{\changes{std}}}
\newcommand{\zqp}{z^\textsc{\changes{std}}}
\newcommand{\lqp}{\lambda^\textsc{\changes{std}}}
\newcommand{\kqp}{\kappa^\textsc{\changes{std}}}
\newcommand{\psoc}{p^\textsc{\changes{cor}}}
\newcommand{\lsoc}{\lambda^\textsc{\changes{cor}}}
\newcommand{\zsoc}{z^\textsc{\changes{cor}}}
\newcommand{\ksoc}{\kappa^\textsc{\changes{cor}}}

\newcommand{\tb}{\textcolor{black}}

\newcommand{\changes}{\textcolor{black}}
\newcommand{\critical}{\textcolor{black}}

\begin{document}

\author{
Xinliang~Dai, 
Junyi~Zhai, 
Yuning~Jiang$^*$, 
Yi~Guo,
Colin N. Jones,  
Veit Hagenmeyer 
\thanks{$^*$corresponding ({\tt yuning.jiang@ieee.org}). 
XD and VH are with Karlsruhe Institute of Technology, JZ is with China University of Petroleum, YJ and CJ are with EPFL, YG is with Empa.}
}



\title{Advancing Distributed AC Optimal Power Flow for Integrated Transmission-Distribution Systems}

\maketitle
\setlength\abovedisplayskip{1pt}
\setlength\belowdisplayskip{1pt}

\begin{abstract}

This paper introduces a distributed operational solution for coordinating \acrfull{itd} systems regarding data privacy.
To tackle the nonconvex challenges of AC \acrfull{opf} problems,
our research proposes an enhanced version of the \acrfull{aladin}. This proposed framework incorporates a second-order correction strategy and convexification, thereby enhancing numerical robustness and computational efficiency. 
The theoretical studies demonstrate that the proposed distributed algorithm operates the \acrshort{itd} systems with a local quadratic convergence guarantee.
Extensive simulations on various ITD configurations highlight the superior performance of our distributed approach in terms of convergence speed, computational efficiency, scalability, and adaptability. 
\end{abstract}

\begin{IEEEkeywords}
\acrfull{itd} systems, AC \acrfull{opf}, distributed nonconvex optimization, second-order correction, \acrfull{aladin}\vspace{-5pt}
\end{IEEEkeywords}

\section{Introduction}


Despite physical connections between transmission and distribution grids, they are traditionally operated separately by transmission system operators (TSOs) and distribution system operators (DSOs), under the assumption that distribution grids can autonomously manage the resultant power flows~\cite{Arthur2020TSODSO}. This operational paradigm is increasingly challenged during the energy transition~\cite{muhlpfordt2021distributed,ZhaiITD2022,bragin2021tso}. The rising integration of 
\acrfull{ders} and other prosumers, who both consume and produce energy, have intensified the interaction between transmission and distribution grids. Challenges\critical{,} such as maintaining voltage within safe limits, managing two-way power flows, and preventing overloads or outages\critical{,} underscore the growing need for adequate coordination of \acrfull{itd} systems to ensure effective and efficient power system operation. 

One strategy for managing \acrshort{itd} systems is to consider them as a unified entity. This contrasts with other coordination strategies discussed in~\cite{Arthur2020TSODSO}, aiming to find an optimal rather than a suboptimal solution. However, such an approach requires the collection of detailed grid data and sensitive customer information by a centralized entity. Centralized approaches are not preferred by system operators or are even forbidden by the respective regulation. For instance, in the United Kingdom, such centralized operation between TSOs and DSOs becomes nearly impossible under the deregulated electricity market environment. 
As an alternative, distributed approaches facilitate independent operation while enabling effective collaboration through limited information sharing. These distributed operation frameworks maintain data privacy and decision-making, spurring significant research into distributed methodologies for ITD system management and coordination~\cite{LiZhengshuo2018,Mohammadi2019,ZhaiJunyialadin,LinChenhui2020,dai2022rapid,bauer2022shapley,dai2023hypergraph}.
\changes{This paper focuses on the AC \acrfull{opf} problems of \acrshort{itd} systems, emphasizing economic efficiency.} Notably, the problem is generally NP-Hard, even for radial power grids~\cite{bienstock2019strong,lehmann2015ac}.
The main challenge in distributed approaches, similar to centralized approaches, is the inherent nonconvex nature resulting from AC power flow equations. 
Previous studies on distributed AC \acrshort{opf} has explored various methodologies, including \acrfull{ocd}~\cite{Hug2009}, \acrfull{app}~\cite{Baldick1999}, \acrfull{dqa}~\cite{Mohammadi2019}, and \acrfull{admm}~\cite{Erseghe2014ADMM,HeXing2020,oh2022distributed,HuangBonan2022}. However, these algorithms lack a convergence guarantee for the original nonconvex problems. Some notable exceptions are the two-level variant of \acrshort{admm}~\cite{sun2021two}, the $l_1$ proximal surrogate Lagrangian method~\cite{bragin2021tso} and the heterogeneous decomposition algorithm~\cite{LiZhengshuo2018}. Nonetheless, these approaches, as first-order algorithms, exhibit slow numerical convergence and limited solution accuracy.
\changes{In response to these challenges, \cite{LinChenhui2020} introduced a two-layer distributed \acrfull{dcc} framework. This framework first solves the convexified subproblems of the distribution grids in the lower layer using the conic relaxation method~\cite{farivar2013branch}, then addresses the nonconvex transmission grid subproblem in the upper layer using second-order curvature information from the distribution subproblems. This approach significantly enhances computational efficiency and provides reliable convergence guarantees for specific network topologies, i.e., star-shaped network configurations with multiple distribution grids connected to a single transmission grid. However, its adaptability is limited in more diverse and realistic network configurations, such as multiple transmission grids, meshed distribution grids, or a combination of multiple grids interconnected in a meshed topology.}

\changes{In contrast, the \acrfull{aladin} algorithm, developed for a broad range of generic nonconvex distributed problems without network topology limitations, offers several advantages over the aforementioned distributed approaches. Utilizing the \acrfull{sqp} strategy, \acrshort{aladin} provides local convergence guarantees with a quadratic convergence rate, without network topology limitations. 
Additionally, \acrshort{aladin} can achieve a global convergence by implementing the globalization strategy~\cite[Alg.~3]{Boris2016}. Crucially, this framework eliminates the need to exchange the original grid data and other data containing customer behaviors, thereby maintaining information privacy. Recently, it has been successfully applied to solve the AC \acrshort{opf} of medium-scale transmission systems~\cite{Engelmann2019} and of AC/DC hybrid  systems~\cite{ZhaiJunyialadin,Meyer2019}, highlighting its capacity to manage distributed problems involving heterogeneous models. However, while standard \acrshort{aladin} shows notable scalability for unconstrained and equality-constrained nonconvex optimization problems~\cite{Boris2016,muhlpfordt2021distributed,dai2022rapid}, it faces scalability challenges in the presence of inequality constraints, as evidenced in studies limited to systems with typically under 300 buses and fewer than half a dozen subgrids.}

\changes{The scalability issue primarily arises from the exponential growth in the number of possible active sets associated with an increase in inequalities, a result of employing the active-set method, leading to combinatorial complexity~\cite[Ch.~15.2]{nocedal2006numerical}. Direct application of conic relaxation, as the \acrshort{dcc} framework~\cite{LinChenhui2020}, results in the weakly active conic constraints, where both constraint residuals and the corresponding Lagrangian multipliers are approaching zero~\cite{tondel2003algorithm}  further complicates the numerical issue associated with the active-set method. Moreover, inaccuracies in the linearization of active constraints, a common issue in \acrshort{sqp}-type methods~\cite[Ch.~18.3]{nocedal2006numerical}, can result in poor initial guesses for subsequent iterations,
posing a significant challenge in the preliminary iterations of augmented-Lagrangian-type algorithms. These issues underscore the need for further refinement to enhance \acrshort{aladin}'s numerical robustness.}

The present paper aims to integrate the strengths of both \acrshort{dcc} and \acrshort{aladin} approaches while simultaneously mitigating their individual limitation. Our focus centers on refining the distributed AC \acrshort{opf} of \acrshort{itd} systems, enhancing the \acrshort{aladin} framework to tackle numerical challenges and boost computational efficiency while preserving its strengths in convergence assurance, accuracy, and adaptability to varied and realistic network structures. Our contributions are as follows:
\begin{enumerate}
    \item \changes{we propose an advanced formulation of distributed AC \acrshort{opf} and employ a new variant of \acrshort{aladin} for solutions. This novel framework is capable of managing more diverse and realistic \acrshort{itd} systems. It effectively handles topologies involving multiple transmission grids, meshed distribution grids, or a combination of multiple grids interconnected in a meshed topology,  thus elevating its adaptability to various network structures.}
    
    \item \changes{ We introduce \acrfull{aladincor}, an refined version of the \acrfull{aladinstd}, to overcome its previously identified numerical limitations. 
    To mitigate the combinatorial challenges posed by the conic relaxation method, we strategically implement relaxation solely in the decoupled step. This approach notably reduces the computational burden on distribution subproblems and simplifies the active-set detection for the relaxed conic constraints.
    Furthermore, we integrate a second-order correction step~\cite{fletcher1982second}~\cite[Ch.~18.3]{nocedal2006numerical} to compensate for the linearization error. This correction is selectively activated during critical iterations marked by significant constraint violations, and we provide rigorous proof affirming the maintenance of the local quadratic convergence rate with this additional correction step.}
   
    \item Numerical investigations are conducted to compare the performance of the proposed \acrshort{aladincor} with the two-layer \acrshort{dcc}~\cite{LinChenhui2020} and the \acrshort{aladinstd} \cite{Boris2016}, demonstrating that the proposed methodology outperforms the others in terms of convergence speed, computational efficiency, solution accuracy, scalability, and adaptability to grid topologies. These results validate that \acrshort{aladincor} effectively combines the advantages of its preprocessors while successfully mitigating their limitations, thus providing a more robust and efficient solution for tackling the complexities of \acrshort{itd} systems.
\end{enumerate}

\section{System Model and Problem Formulation}\label{sec::formulation}
This section presents the system model of an \acrshort{itd} system and then formulate the AC \acrshort{opf} of the \acrshort{itd} system in a distributed form with affine consensus constraints.
\subsection{System Model of ITD Systems}
We describe a \acrshort{itd} system by a tuple $\mathcal{S}=(\mathcal{R},\;\mathcal{N},\;\mathcal{L})$, where $\mathcal{R} = \mathcal{R}_T \bigcup \mathcal{R}_D$ represents the set of all subgrids, or so-called regions, $\mathcal{R}_T=\{T_1,\dots,T_m\}$ the set of $m$ transmission grids, and $\mathcal{R}_D=\{D_1,\dots,D_n\}$ the set of $n$ radial distribution grids, $\mathcal{N}$ denotes the set of all buses, $\mathcal{L}$ the set of all branches. In a specific region $\ell\in\mathcal{R}$, $\mathcal{N}_\ell$ denotes the set of buses and $\mathcal{L}_\ell$ denotes the set of branches in the region $\ell$. Additionally, the tie-line connected to neighboring regions is also included in the set of branches in the region $\ell$, i.e., $\mathcal{L}_\ell^\textrm{tie}\subseteq\mathcal{L}_\ell$.

We adopt \acrfull{bim} for transmission systems and \acrfull{bfm} for radial distribution systems~\cite{baran1989optimal1,baran1989optimal2}, setting the stage for the conic relaxation within the \acrshort{aladin}-type algorithm. Our proposed methodology diverges from~\cite{LinChenhui2020} in its capability to handle multiple transmission systems. This allows us to handle meshed distribution grids analogously to transmission grids, enhancing the adaptability to diverse grid topologies.

In the present paper, we consider both control and state variables in each model as decision variables. For a transmission grid denoted as $T_\tau\in\mathcal{R}_T$, the decision variables include voltage magnitude $v_i$, voltage angle $\theta_i$, active power generation $p_i^g$ and reactive power generation $q_i^g$ for all bus $i\in\mathcal{N}_{T_\tau}$. Similar to a distribution grid denoted as $D_\delta\in\mathcal{R}_D$, the decision variables encompass squared voltage magnitude $u_i = v_i^2$, active power generation $p_i^g$ and reactive power generation $q_i^g$ for all bus $i\in\mathcal{N}_{D_\delta}$, as well as squared current magnitude $l_{ij}$, active and reactive power flow $p_{ij}$, $q_{ij}$ for each transmission line $(i,j)\in\mathcal{L}_{D_\delta}$. \changes{We collected the decision variables in transmission systems in the vector $x_{T_\tau}$, and the decision variables in distribution systems in the vector, $x_{D_\delta}$.} 

\begin{remark}
    In the present paper, we refer to the power supplied by generators in transmission systems and \acrshort{ders} in distribution systems collectively as generation power. Notably, the power supplied by \acrshort{ders} may assume negative values in certain scenarios.
\end{remark}

\subsection{Network Decomposition and Consensus Constraints}\label{sec::formulation::consensus}

In this paper, we are set to propose a distributed algorithm tailored for real-time dispatch for \acrshort{itd} systems. Before we present the dispatch problem formulation, we here firstly introduce a network decomposition and associated consensus constraints based on the network configuration. 

Ensuring numerical equivalence between distributed and centralized \changes{approaches} for \acrshort{itd} systems, effective system decomposition and the proper configuration of consensus constraints are crucial.
\changes{Our approach employs a component-sharing strategy among neighboring regions. 
This strategy is designed to maintain the physical consistency of the \acrshort{itd} system without introducing additional variables or modeling assumptions. This approach contrasts with methods like those in~\cite{Engelmann2019}, which suggest cutting transmission lines and adding auxiliary buses. By avoiding these additional modeling complexities, the adoption of this component-sharing strategy ensures that no structural numerical errors are introduced into the system model~\cite{muhlpfordt2021distributed,Erseghe2014ADMM}, and thus forms the cornerstone of our proposed distributed approaches for effective real-time dispatch in \acrshort{itd} systems.}

\begin{figure}[htbp!]
    \centering
    \subfloat[Coupled \acrshort{itd} system]{\includegraphics[width=0.22\textwidth]{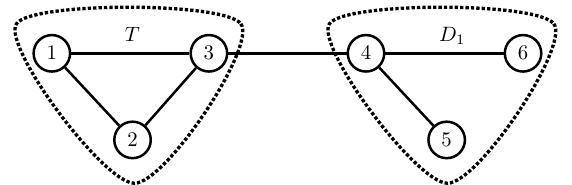}}\\[-0.2cm]
    \subfloat[Transmission $T_1$ (\acrshort{bim})]{\includegraphics[width=0.18\textwidth]{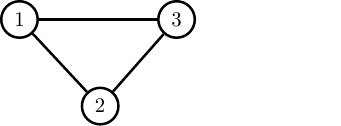}}\qquad
    \subfloat[Distribution $D_1$ (\acrshort{bfm})]{\includegraphics[width=0.18\textwidth]{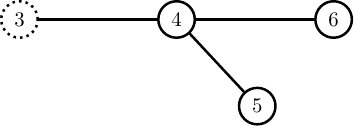}} 
    \caption{Decomposition between transmission and distribution\label{fig::itd}}
\end{figure}

\changes{Facilitating this component-sharing strategy entails defining \textit{core} and \textit{copy} buses for each region. For a given region $\ell\in\mathcal{R}$, we define the \textit{core} buses as those within its own domain, while \textit{copy} buses are those replicated from neighboring regions. }
Consequently, the bus set for the given region $\ell$ is represented as $\mathcal{N}_\ell =\mathcal{N}^\textrm{core}_\ell\bigcup\mathcal{N}^\textrm{copy}_\ell$. Furthermore, considering the two different models within \acrshort{itd} systems, i.e., \acrshort{bim} and \acrshort{bfm},  connecting tie lines between two subgrids are categorized into Transmission-Distribution (T-D), Transmission-Transmission (T-T), or Distribution-Distribution (D-D) types. The corresponding branch sets are denoted as $\mathcal{L}^{TD}$,  $\mathcal{L}^{TT}$ and $\mathcal{L}^{TT}$ respectively. 
\changes{For illustration, we present a two-region system with 6 buses in~\autoref{fig::itd}(a), demonstrating the practical application of our component-sharing strategy in a \acrshort{itd} system configuration.} 

\vspace{-2pt}
\begin{figure}[htbp!]
    \centering
     \subfloat[Transmission $T_1$ (\acrshort{bim})]{\includegraphics[width=0.18\textwidth]{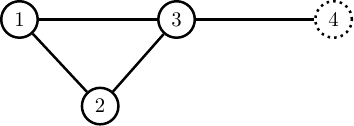}}\qquad
    \subfloat[Transmission $T_2$ (\acrshort{bim})]{\includegraphics[width=0.18\textwidth]{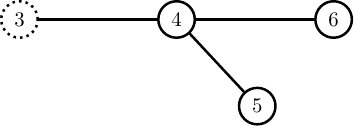}}    
    \caption{Decomposition between two transmissions\label{fig::itt}}
\end{figure}\vspace{-2pt}

In the case of T-D connections, depicted in~\autoref{fig::itd}(b)(c), the distribution grid $D_1$ inclusdes \textit{core} buses $\{4,5,6\}$ and a \textit{copy} bus $\{3\}$  replicated from the neighboring transmission grid, whereas the transmission grid $T$ has no \textit{copy} bus. The coupling variables here are the squared voltage magnitude at bus~$3$, active and reactive power flow along the connecting tie-line~$(3,4)$, denoted as $
\{u_3,p_{34},q_{34}\}$.  The consensus constraints for setup are expressed as
\begin{equation*}\label{eq::example::consensus::equations}
    u^{\textrm{copy}}_3 = u^{\textrm{core}}_3,\;p^{\textrm{copy}}_{34} = p^{\textrm{core}}_{34},\;q^{\textrm{copy}}_{34} = q^{\textrm{core}}_{34}.
\end{equation*}
A similar approach is adopted for D-D connections, with an extra step to identify a primary region.

In T-T connections, as illustrated in~\autoref{fig::itt}(a) and (b), the coupling variables include voltage magnitudes and angles at the buses~$\{3,4\}$, denoted as $\{v_3,v_4,\theta_3,\theta_4\}$. Different from~\autoref{fig::itd}(b)(c), both transmission grids include an additional \textit{copy} bus replicated. The resulting consensus constraints between two transmissions are
\begin{equation*}\label{eq::example::consensus::equations::tt}
    v^{\textrm{copy}}_3 = v^{\textrm{core}}_3,\;v^{\textrm{copy}}_4 = v^{\textrm{core}}_4,\;\theta^{\textrm{copy}}_3 = \theta^{\textrm{core}}_3,\;\theta^{\textrm{copy}}_4 = \theta^{\textrm{core}}_4.
\end{equation*}

In conclusion, the consensus constraints in a specific \acrshort{itd} system are linear and can be formulated into an affinely coupled form
\begin{equation}\label{eq::example::consensus::matrix}
    \sum_{\ell\in\mathcal{R}}A_\ell x_\ell= Ax=b
\end{equation}
with consensus matrix $A = \left[A_{T_1},\cdots,A_{T_m},A_{D_1},\cdots,A_{D_n}\right]$ and the decision variables $x = [x_{T_1},\cdots,x_{T_m},x_{D_1},\cdots,x_{D_n}]$. Note that $b$ is a zero vector in the proposed model.

The integrating of \textit{core} and \textit{copy} buses allows us to formulate self-contained AC \acrshort{opf} subproblems for individual regions. Coupled with the consensus constraints designed to assure the numerical consistency of the coupling variables, it guarantees that solutions obtained through both distributed and centralized approaches are equivalent.
\begin{remark}\label{rem::redundant}
\changes{
In distributed approaches utilizing the component-sharing strategy, maintaining the \acrfull{licq} is crucial.
To ensure this, nodal constraints---such as nodal power balance and limits on nodal decision variables---are applied solely to the core buses in each region. 
This is essential because the numerical equivalence between core and copy variables has already been established by consensus constraints.
Extending these nodal constraints to the copy buses, which are essentially replicas from neighboring regions, would result in redundant constraints in the optimization problem and consequently violate the \acrshort{licq}.
}
\end{remark}
\subsection{Model of Transmission Systems}
In the present paper, we model transmission systems using the standard \acrfull{bim}. For the specific region $\ell\in\mathcal{R}_T$, we let $Y= G+\text{j}B$ denote the complex bus admittance matrix of the transmission system, where $G,B\in\mathbb{R}^{|\mathcal{N}_\ell|\times|\mathcal{N}_\ell|}$. Let $p_i^g$ and $q^g_i$ (resp. $p^l_i$ and $q^l_i$) denote the active and reactive power injection by generator (resp. load) at bus $i$. \changes{Consequently, active and reactive power injections at \textit{core} bus $i\in\mathcal{N}_\ell^\text{core}$ can be expressed as}
\begin{align}
        p_{i} \;= \;p_i^g-p_i^l - p_i^{TD},\;q_{i} \;= \; q_i^g-q_i^l - q_i^{TD},
\end{align}
\changes{where $p_i^{TD}$ and $q_i^{TD}$ summarized active and reactive power delivered to distribution systems via T-D connections at bus $i\in\mathcal{R}_\ell$, i.e.,}
\begin{equation}
    p_i^{TD} = \sum_{\substack{(i,j)\in\mathcal{L}^\text{TD}}} p_{ij},\quad q_i^{TD} = \sum_{\substack{(i,j)\in\mathcal{L}^\text{TD}}} q_{ij}.
\end{equation}
\changes{If no distribution network is connected to bus $i$, then $p_i^{TD}=q_i^{TD}=0$, same as the load and generator.}

The mathematical model of the transmission system can be written as
\begin{subequations}\label{eq::bim}
    \begin{align}
        p_i &= \, v_i \sum_{j\in\mathcal{N}_\ell} v_j \left( G_{ij} \cos\theta_{ij} + B_{ij} \sin\theta_{ij} \right),  \; i\in\mathcal{N}_\ell^\text{core}\label{eq::bim::balance::p}\\
        q_i &=  \,v_i \sum_{j\in\mathcal{N}_\ell} v_j \left( G_{ij} \sin\theta_{ij} - B_{ij} \cos\theta_{ij} \right),  \; i\in\mathcal{N}_\ell^\text{core}\label{eq::bim::balance::q}\\
        p_{ij}^2 &  +\,  q_{ij}^2 \leq \; \overline{s}_{ij}^2, \qquad \qquad \qquad \qquad\quad  \; \; \, (i,j)\in \mathcal{L}_\ell\label{eq::bim::limit::line}\\
        \underline v_i & \leq v_i \leq \overline v_i,\,
        \underline p_i^g \leq p_i^g \leq \overline q_i^g,\,
        \underline q_i^g \leq q_i^g \leq \overline q_i^g,\,i\in\mathcal{N}_\ell^\text{core}\label{eq::bim::limit::variables}
    \end{align}
\text{with active and reactive power flow on branch $(i,j)\in\mathcal{L}_\ell$}
    \begin{align}
        \quad p_{ij}=&\; v_i^2 \,G_{ij}-v_i v_j\left(G_{ij}\cos \theta_{ij}+B_{ij}\sin \theta_{ij}\right),\\
        q_{ij}=&-v_i^2 \, G_{ij} - v_i v_j\left(G_{ij}\sin \theta_{ij}-B_{ij}\cos \theta_{ij}\right).
    \end{align}
\end{subequations}
where $\overline{s}_{ij}$ denote apparent power limit on the branch $(i,j),$ and $\underline v_i$, $\overline v_i$, $\underline p_i^g$, $\overline q_i^g$, $\underline q_i^g$, and $\overline q_i^g$ denote the upper and lower bounds for the corresponding decision variables.
The \acrshort{bim} thus encompasses the active and reactive nodal power balances~\eqref{eq::bim::balance::p}\eqref{eq::bim::balance::q}, apparent power limit on transmission lines~\eqref{eq::bim::limit::line} and the bounds on the voltage magnitudes and power generations~\eqref{eq::bim::limit::variables}. 

\begin{remark}\label{rem::connection}
    
    \changes{In the transmission system model~\eqref{eq::bim}, a \textit{copy} bus is always replicated from an adjacent transmission because connecting to a distribution system does not require introducing a \textit{copy} bus. Moreover, the right-hand sides of~\eqref{eq::bim::balance::p}\eqref{eq::bim::balance::q} aggregate all the active and reactive power flows from the adjacent buses to bus $i$, including all the adjacent bus $j\in\mathcal{N}_\ell = \mathcal{N}_\ell^\text{core} \bigcup \mathcal{N}_\ell^\text{copy}$.  Consequently, power flows via T-T connections are appropriately taken into consideration in the nodal power balance~\eqref{eq::bim::balance::p}\eqref{eq::bim::balance::q}; for more detail, we reference to the discussion on distributed formulation in~\cite{muhlpfordt2021distributed}.}

\end{remark}

\subsection{Model of Distribution Systems}
In our model, the remaining radial distribution network is represented using \acrfull{bfm} introduced in~\cite{baran1989optimal1,baran1989optimal2}.  For a specific region $\ell\in\mathcal{R}_D$, let $l_{ij}$, $r_{ij}$ and $\chi_{ij}$ denote the squared current magnitude, the resistance and reactance, respectively, and $\overline{l}_{ij}$ denotes as current limit on the branch $(i,j)$. Squared voltage magnitude at bus $i$ and its upper and lower bounds are given by $u_i$, $\overline{u}_i$ and $\underline{u}_i$. The mathematical model can be written as
\begin{subequations}\label{eq::bfm}
    \begin{align}
         u_j =& \,u_i - 2 (r_{ij}\,p_{ij} + \chi_{ij}\,q_{ij}) \notag \\
        &\qquad\qquad+(r_{ij}^2+\chi_{ij}^2) \,l_{ij}, \quad\forall (i,j)\in\mathcal{L}_\ell\label{eq::bfm::balance}\\
        p_j   =&\, \sum_{k\in\mathcal{N}_\ell} p_{jk} - \sum_{i\in\mathcal{N}_\ell} (p_{ij} -r_{ij} \, l_{ij}), \; \forall j \in\mathcal{N}_\ell^\text{core}\label{eq::bfm::act::balance}\\
        q_j   =&\, \sum_{k\in\mathcal{N}_\ell} q_{jk} - \sum_{i\in\mathcal{N}_\ell} (q_{ij} -\chi_{ij} \, l_{ij}), \; \forall j \in\mathcal{N}_\ell^\text{core}\label{eq::bfm::react::balance}\\
         l_{ij} =&\,  \frac{p_{ij}^2+q_{ij}^2}{u_i},\quad \forall (i,j)\in\mathcal{L}_\ell \label{eq::bfm::nonconvex}\\
         l_{ij} \leq&\, \overline{l}_{ij},\quad \forall (i,j)\in\mathcal{L}_\ell\label{eq::bfm::branch::limit}\\       
        \underline u_i \leq & \,u_i \leq  \overline u_i,\,
        \underline p_i^g \leq p_i^g \leq \overline q_i^g,\,
        \underline q_i^g \leq q_i^g \leq \overline q_i^g,\,\forall i \in\mathcal{N}_\ell^\text{core}\label{eq::bfm::bounds}
    \end{align}
\end{subequations}
where $p_j = p_j^g-p_j^l$, $q_j = q_j^g-q_j^l$.
Equations \eqref{eq::bfm::balance}-\eqref{eq::bfm::nonconvex} are DistFlow constraints, and \eqref{eq::bfm::branch::limit} denotes the current limits on each branch $(i,j)\in\mathcal{L}_\ell$. Voltage magnitude and power generation limits at each bus $i\in\mathcal{N}_\ell$ are constrained by respective upper and lower bounds~\eqref{eq::bfm::bounds}. \changes{Similar to the analysis in Remark~\ref{rem::connection}, power flows via T-D and D-D connections are taken into consideration in the nodal power balance~\eqref{eq::bfm::act::balance}\eqref{eq::bfm::react::balance}, since they aggregate power flows from all adjacent buses, including \textit{copy} buses.}

Notice that the feasible set of \eqref{eq::bfm::nonconvex} is still nonconvex due to the quadratic equality constraint~\eqref{eq::bfm::nonconvex}. For computational efficiency, a classic method~\cite{farivar2013branch} is to apply an exact conic relaxation due to radial network topology, resulting in a convex feasible set. However, directly applying the conic relaxation to the model of distribution systems, will introduce weakly active inequality constraint. \changes{These constraints pose a challenge as both the constraint residuals and the corresponding Lagrangian multipliers typically approach zero~\cite{tondel2003algorithm}. This scenario complicates the numerical challenges associated with the active-set method~\cite[Ch.~15.2]{nocedal2006numerical}. To tackle this challenge effectively in our distributed approach, we apply the relaxation specifically while solving the decoupled subproblems. Such a customized implementation allows us to distinguish the weakly active conic constraints from other inequality constraints during the active-set detection and treat them as equality constraints instead, thereby streamlining the overall problem-solving process.}

\subsection{Objective}

We consider a general cost function that is applicable to both transmission and distribution grids, accommodating diverse operation objectives of different subgrids. The cost for any given specific region $\ell\in\mathcal{R}$ encompassing operation costs of generators in transmission grid or \acrshort{ders} in distribution, as well as the penalties on line losses, can be written as
\begin{align}\label{eq::objective::global}
        f_\ell (x_\ell) = c_\ell^\text{gen} \sum_{i\in\mathcal{N}_\ell} \left\{a_{i2} \left(p^{g}_i\right)^2+a_{i1} p^{g}_{i} + a_{i0}\right\}\notag\\
        + c_\ell^\text{loss} \sum_{(i,j)\in\mathcal{L}_\ell} r_{ij} l_{ij},
\end{align}
where $a_{i2}$, $a_{i1}$, and $a_{i0}$ denote the polynomial coefficients of operation cost of generator or \acrshort{ders} at bus $i$. The weighted coefficients $c_\ell^\text{gen}$, $c_\ell^\text{loss}$ correspond to the operation costs and power loss penalties, respectively.

The objective function formulation~\eqref{eq::objective::global} allows  TSOs and DSOs to independently tune the coefficients $c_\ell^\text{gen}$ and $c_\ell^\text{loss}$ according to their distinct operational needs.
Such a framework ensures that a wide range of operational tasks can be effectively represented and managed within the proposed model.


\color{black}
\subsection{Distributed Formulation}
Based on the discussion given above in the present section~\ref{sec::formulation}, the AC \acrshort{opf} of \acrshort{itd} systems can be formulated in the standard affinely coupled distributed form
\begin{subequations}\label{eq::formulation}
\begin{align}
\min_{x}\quad&f(x):=\sum_{\ell\in\mathcal{R}} f_\ell(x_\ell)\\\label{eq:affine1}
\textrm{s.t.}  \quad  &\sum_{\ell\in\mathcal{R}} A_\ell x_\ell =b\quad\;\;\mid\lambda\\\label{ineq::nonlinear}
&h_\ell(x_\ell)\leq 0\qquad\quad\mid\kappa_\ell,\;\ell\in\mathcal{R},
\end{align}
\end{subequations}
where \eqref{ineq::nonlinear} summarizes all local constraints~\eqref{eq::bim}---\eqref{eq::bfm} for transmission and distribution systems, and  $\lambda$, $\kappa_{\ell}$ denote the dual variables (Lagrangian multipliers) associated with the constraints~\eqref{eq:affine1} and~\eqref{ineq::nonlinear}, respectively. 

Note that the constraint~\eqref{ineq::nonlinear} includes detailed grid topology and other sensitive customer data, such as consumer behaviors. Centralized approaches, requiring collection and access to all these data in a centralized entity, are not preferred due to privacy and practical concerns. Distributed approaches, as alternatives, ensure data privacy by allowing each region access to its own dataset, with limited and sometimes encrypted information exchange.

Despite this, the distributed problem~\eqref{eq::formulation} remains challenging due to the inherent nonconvexity of the transmission and distribution models. This paper, therefore, focuses on exploring the numerical optimization algorithms in this distributed context. The next section will present our proposed distributed optimization algorithm, offering insight into its convergence properties and implementation details.   


\section{Distributed Optimization Framework}\label{sec::algorithm}

In this section, we present a novel variant of \acrfull{aladin} algorithm for solving the distributed nonlinear and nonconvex AC \acrshort{opf} problem of \acrshort{itd} systems with convergence analysis and implementation details.

\subsection{ALADIN with Second-Order Correction}\label{sec::alg::soc}

\changes{The \acrfull{aladinstd}, tailored for generic distributed nonconvex optimization problems~\cite{Boris2016}, stands out by integrating \acrfull{admm} with the \acrfull{sqp} framework.
This integration enables it to achieve locally quadratic convergence rates in a distributed setting without limitations on network topology, marking a significant advancement over other distributed approaches. However, inaccuracies in the linearization of active constraints, a common issue in \acrshort{sqp}-type methods, can result in poor initial guesses for subsequent iterations. Such inaccuracies can lead to suboptimal initial estimates for the subsequent iterations, posing a significant challenge in the preliminary iterations of augmented-Lagrangian-type algorithms. }

\changes{To mitigate the issue, we proposed a novel variant of \acrshort{aladinstd} incorporates the second-order correction method to compensate for the linearization error~\cite{fletcher1982second}~\cite[Ch.~18.3]{nocedal2006numerical}. The correction step is selectively applied, not in every iteration but specifically when a trial step is significantly impacted by linearization errors. For effective decision-making, we utilize an $L_1$ penalty function $\Phi(x)$ as a merit function to assess the quality of a given trial step $x$}
\begin{equation}\label{eq::merit}
    \Phi(x) = \sum_{\ell\in\mathcal R}f_\ell(x_\ell) +\zeta \norm{\sum_{\ell\in\mathcal R}A_\ell x_\ell -b}_1 +  \xi\cdot\psi(h(x)),
\end{equation}
\changes{where $\psi (h) = \sum_i\max\{0, [h]_i\}$ measures the overall constraint violation. The positive penalty parameters $\zeta,\;\xi$ are assumed to be sufficiently large such that $\Phi$ is an exact penalty function for the problem~\eqref{eq::formulation}. The correction step is then executed if the merit function $\Phi(x)$ increases coupled with considerable constraint violations $\psi (h(x))$.}
\begin{definition}[\cite{nocedal2006numerical}]
    A penalty function is \textit{exact} if a single minimization with respect to $x$ can yield the exact solution of the original constrained optimization problem
\end{definition}
\begin{algorithm}[htbp!]
    \footnotesize
    \caption{Full-Step \acrshort{aladincor}\label{alg::aladin}}
    \textbf{Initialization:} define $L_1$ penalty function $\Phi(x)$~\eqref{eq::merit}\\
    \textbf{Input}: initial primal and dual points $(z,\;\lambda)$, positive penalty parameters\;$\rho$,\;$\mu$ and scaling symmetric matrices $\Sigma_\ell\succ 0$\\
    \textbf{Repeat:}
    \begin{enumerate}[leftmargin=10pt]
    \item solve the following decoupled \acrshort{nlp}s for all \tb{$\ell\in\mathcal{R}$} \label{alg::aladin::s1}
    \begin{subequations}
    	\label{alg::aladin::nlp}
    	\begin{align}
    		\min_{x_\ell}\quad &f_\ell(x_\ell)+\lambda^\top A_\ell x_\ell+
    		\frac{\rho}{2}\norm{x_\ell-z_\ell}^2_{\Sigma_\ell}\\
    		\textrm{s.t.}\quad &h_\ell(x_\ell)\leq 0\qquad\mid\kappa_\ell
    	\end{align}
    \end{subequations}
    \item compute the Jacobian matrix $J_\ell$ of active constraints $h^\textrm{act}_\ell$ at the local solution $x_\ell$ by \label{alg::aladin::s2}
    \begin{equation}
        J_\ell = \nabla h^\textrm{act}_\ell(x_\ell)\label{eq::Jacobian},    
    \end{equation}
    and gradient $g_\ell=\nabla f_\ell(x_\ell)$, choose Hessian approximation
    \begin{equation}\label{eq::Hessian}
    	\tb{H_\ell\approx\nabla^2\left\{f_\ell(x_\ell)+\kappa_\ell^\top h_\ell(x_\ell)\right\}\succ 0},
    \end{equation}
    \item terminate if $\norm{Ax - b}_2 \leq \epsilon$ and $\norm{\Sigma(x-z)}_2 \leq \epsilon$ are satisfied. \label{alg::aladin::s3}
    \item obtain $(\zqp=x+\pqp,\lqp)$ by solving coupled \acrshort{qp} \label{alg::aladin::s4}
    \begin{subequations}\label{alg::aladin::qp}
    \begin{align}
    	\min_{\pqp,s}\;\;& \sum_{\ell\in\mathcal{R}}\left\{\frac{1}{2} \left(\pqp_\ell\right)^\top H_\ell \; \pqp_\ell + g_\ell^\top \pqp_\ell\right\}
    	+\lambda^\top\;s + \frac{\mu}{2} \norm{s}^2_2\\\label{eq::slack::consensus}
    	\textrm{s.t.}\;\;\;& \quad  \sum_{\ell\in\mathcal{R}} A_\ell (x_\ell+ \pqp_\ell) = b + s \quad \mid\lqp\\\label{eq::active}
    	&  \quad J_\ell  \;\pqp_\ell = 0,\;\;\ell \in \mathcal{R}
    \end{align}
    \end{subequations}
    \item if the value of $\Phi$ increases due to the intolerant violation of active constraint~$h^\textrm{act}$ at the new iterate $\zqp$, compute $(\zsoc=x+\psoc, \lsoc)$ by\label{alg::aladin::s5}
    \begin{align}  \label{eq::soc::lp}
    \begin{bmatrix}
        \psoc\\ \lsoc
    \end{bmatrix} 
     = \begin{bmatrix}
        \pqp\\ \lqp
    \end{bmatrix} -
    \begin{bmatrix}
         I\\ \mu A
    \end{bmatrix}\cdot M \, J^\top \cdot \left(J \, M \, J^\top\right)^{-1} \cdot r
    \end{align}
    with $r = h^\textrm{act}(x+\pqp)$ and $ M = \left(H+\mu A^\top A \right)^{-1}$.
	\item update the primal and the dual variables with full step  \label{alg::aladin::s6}
    \begin{equation}
        (z^{+},\; \lambda^{+}) = \begin{cases}
        (\zsoc, \;\lsoc), & \textrm{step 5 executed},\\[0.12cm]
        (\zqp, \;\lqp), & \textrm{otherwise}.
        \end{cases}
    \end{equation}
    \end{enumerate}
\end{algorithm}\vspace{-2pt}
The adapted variant, socalled \acrfull{aladincor}, is presented in Algorithm \ref{alg::aladin} for solving~\eqref{eq::formulation}. In step~\ref{alg::aladin::s1}, the original decoupled subproblems are reguarized to the decoupled \acrshort{nlp}~\eqref{alg::aladin::nlp} by introducing the augmented Lagrangian method, where $\rho$ is the penalty parameter and $\Sigma_\ell\in\mathbb{R}^{x_\ell \times x_\ell}$ is the scaling matrix for the proximal term. A practical strategy to update $\rho$ for distributed AC \acrshort{opf} can be found in~\cite{Engelmann2019}. 

In Step~\ref{alg::aladin::s2}, all curvature information, i.e., Jacobian $J_\ell$ of active constraints $h^\textrm{act}_\ell$, gradient of local objective $g_\ell$, and Hessian approximation $H_\ell$ are computed. \changes{The active constraint $h^\textrm{act}_\ell(x_\ell)$ at the current iteration encompasses the inequality constraints for all $i \in\mathcal{S}_\ell=\{i\;|\; [h_{\ell}(x_{\ell})]_i=0\}.$}
Note that both steps can be executed in parallel.

After the parallelizable steps, the termination condition is checked by the coordinator. The \acrshort{aladin} algorithm will be terminated if \changes{the primal and dual conditions are satisfied, i.e.,} the primal and the dual residuals are smaller than the predefined tolerance $\epsilon$
\begin{align}
    \norm{Ax - b}_2 \leq \epsilon\;\textrm{and}\;\norm{\Sigma(x-z)}_2 \leq \epsilon\label{eq::dual},
\end{align}
where $\Sigma$ is a block diagonal matrix consists of scaling matrix $\Sigma_\ell$ for all $\ell\in\mathcal{R}$. It indicates that the local solution $x_\ell$ satisfied the first-order optimality condition of the original problem~\eqref{eq::formulation} under the tolerable error of $\mathcal{O}(\epsilon)$, i.e.,
\begin{equation}\label{eq::kkt}
    \norm{\nabla\left\{f_\ell(x_\ell)+\kappa_\ell^\top h_\ell(x_\ell) \right\}+ A_\ell^\top \lambda}_2 =\mathcal{O}(\epsilon).
\end{equation}
\vspace{-0.5cm}
\begin{remark}
    Practically, the dual condition~\eqref{eq::dual} is sufficient to ensure the small violation of the condition~\eqref{eq::kkt}, when the predefined tolerance $\epsilon$ is small enough~\cite{Houska2021}. 
\end{remark}
In the coupled Step~\ref{alg::aladin::s4}, a quadratic approximation of the original problem~\eqref{eq::formulation} is established based on the curvature at local solution $x_\ell$ computed in Step~\ref{alg::aladin::s2}. An additional slack variable $s$ is introduced to avoid the infeasibility of the approximated problem due to the consensus constraint~\eqref{eq::slack::consensus}, and in order to increase numerical robustness. Like the \acrshort{sqp}, the active constraints are linearized and summarized in~\eqref{eq::active}.
\changes{ For \acrshort{aladinstd}, the trial step $\pqp$ would be accepted and the primal and dual variables are updated as $(\zqp=x+\pqp, \lqp)$ }

\begin{remark}
No detailed grid data or any private data related to customer behavior is required in the coupled Step~\ref{alg::aladin::s4}, but the curvature information, including gradient, Jacobian, and approximated Hessian of the local problems~\eqref{alg::aladin::nlp}.  
\end{remark}

\changes{
When there is an increase in the $L_1$ merit function $\Phi(\zqp)$ accompanied by a significant violation of the active constraint $\psi(h^\textrm{act}(\zqp))$, it indicates that the trial step $\pqp$ may be substantially affected by linearization errors. 
In such scenarios, an additional second-order correction step~\ref{alg::aladin::s5} is employed to generate a corrected step $\psoc$ to compensate for the linearization error.
}

Following~\cite[Ch.~18.3]{nocedal2006numerical}, the coupled \acrshort{qp} subproblems~\eqref{alg::aladin::qp} are resolved, with the linearized active constraint~\eqref{eq::active} replaced by 
\begin{equation}
    \quad J_\ell \; \psoc_\ell + r_\ell = 0,\;\;\ell \in \mathcal{R},
\end{equation}
where the residual vector $r_\ell$ computed by
\begin{equation}
    r_{\ell} = h^\textrm{act}_{\ell}(x_{\ell}+\pqp_\ell) - J_{\ell} \pqp_\ell = h^\textrm{act}_{\ell}(x_{\ell}+\pqp_\ell).
\end{equation}
Here, the term $J_{\ell}\cdot \pqp_\ell$ can be neglected because the trial step $\pqp_\ell$ already satisfies the constraints~\eqref{eq::active}. The resulting second-order correction subproblem can be written as
\begin{subequations}\label{alg::aladin::compensated::qp}
\begin{align}\notag
	\min_{\psoc,s}&\;\;\; \sum_{\ell\in\mathcal{R}}\left\{\frac{1}{2} \left(\psoc_\ell\right)^\top\;H_\ell \;\psoc_\ell + g_\ell^\top \psoc_\ell\right\}\\&\qquad \qquad\qquad\qquad\qquad\quad  +\lambda^\top s + \frac{\mu}{2} \norm{s}^2_2\\
	\textrm{s.t.}\;\;  & \quad  \sum_{\ell\in\mathcal{R}} A_\ell (x_\ell+ \psoc) = b + s \quad \mid\lsoc\\
    &  \quad J_\ell \; \psoc_{\ell}+ r_\ell= 0, \;\;\ell \in \mathcal{R}.\label{eq::soc::linearize::constraint}
\end{align}
\end{subequations}
Fortunately, the compensated step $(\psoc,\lsoc)$ can be computed analytically by~\eqref{eq::soc::lp}, in which all matrix inverses and the factorization for solving~\eqref{alg::aladin::qp} can be reused, and no update of curvature information is required. More details will be discussed in section~\ref{sec::soc}.

Local convergence of the proposed \acrshort{aladincor} algorithm is guaranteed, and its analysis will be provided in the next section, while global convergence can be achieved if the additional globalization strategy~\cite[Alogrithm~3]{Boris2016} is implemented. 
\begin{remark}[(Globalization of Algorithm~\ref{alg::aladin})]\label{remark::globalization}
To enforce the global convergence of Algorithm~\ref{alg::aladin} such that it converges to a local minimizer of~\eqref{eq::formulation}, the primal-dual iterate $(z,\lambda)$ is updated by  
    \begin{subequations}~\label{eq::globalization}
    	\begin{align}
    		z^{+} =& z +\alpha_1(x-z) + \alpha_2 \pqp,\\
    		\lambda^{+} =& \lambda + \alpha_3 (\lqp-\lambda).
    	\end{align}
    \end{subequations}
\changes{If the second-order correction step is executed, the trial step $(\pqp,\lqp)$ in~\eqref{eq::globalization} should be replaced by the corrected step $(\psoc,\lsoc)$.} Furthermore, the line search scheme~\cite[Algorithm~3]{Boris2016} can be used to calculate the step sizes $\alpha_1$, $\alpha_2$ and $\alpha_3$.
\end{remark}
\begin{remark}[(Initialization of Algorithm~\ref{alg::aladin})]\label{remark::flat}
    \acrshort{opf} problems are usually initialized with a flat start, where all voltage angles are set to zero and all voltage magnitudes are set to 1.0 p.u.~\cite{frank2016introduction}. Besides, the dual variables of consensus constraints are set to zero for distributed \acrshort{opf}.
    For this initialization strategy, it has been demonstrated numerically that it can provide a good initial guess in practice~\cite{Engelmann2019,ZhaiJunyialadin}. Hence, we focus on the full-step version of \acrshort{aladin}, i.e., $\alpha_1=\alpha_2 = \alpha_3=1$, in the present paper.
\end{remark}

\subsection{Numerical Implementation}\label{sec::alg::practice}
Some modifications are introduced to improve computation complexity. One is the conic relaxation of distribution subproblems in the \acrshort{bfm}, as discussed above. Another is the second-order correction to compensate for the error in the linearization of the active constraints~\eqref{eq::active} in the coupled \acrshort{qp} step~\ref{alg::aladin::s4}. 

\subsubsection{Second-Order Conic Relaxation}
The feasible set of the AC \acrshort{opf} problem for the distribution grid defined by~\eqref{eq::bfm::nonconvex}-\eqref{eq::bfm::bounds} is still nonconvex due to the quadratic equalities~\eqref{eq::bfm::nonconvex}. To further reduce the computational burden and computing time of the decoupled \acrshort{nlp} for distribution grids, we can reformulate the corresponding \acrshort{nlp} as a conic optimization problem by replacing equality constraints \eqref{eq::bfm::nonconvex} with inequality constraints
\begin{equation}\label{eq::bfm::soc}
    l_{ij} \geq \frac{p_{ij}^2+q_{ij}^2}{u_i},\quad \forall (i,j)\in\mathcal{L}_\ell,
\end{equation}

\begin{remark}[\cite{farivar2013branch}]
    The conic relaxation of the \acrshort{opf} problem of a radial grid in the branch flow model is exact if the objective function is convex and strictly increasing in line loss.
\end{remark}

This theorem implies that the solution to the conic relaxed problem aligns with that of the original problem~\eqref{eq::formulation}. Specifically, it indicates that the relaxed inequality constraints are always active or at least weakly active, at the optimal solution, i.e., they form part of the optimal active set:
\begin{equation}
    i\in\mathcal{S}^* = \{i\;|\;[h(x^*)]_i = 0\}.
\end{equation}
\changes{When augmented Lagrangian methods are applied, it's observed that most of the conic residuals stay within acceptable tolerances. For those residuals that initially exhibit small deviations in the preliminary iterations, rapid convergence to zero is typically observed as iterations progress. Consequently, the need for active-set detection for these relaxed conic constraints can be eliminated.}

\subsubsection{Second-Order Correction Step}\label{sec::soc} Based on the solution $(\pqp,\lqp)$ of~\eqref{alg::aladin::qp}, the correction step $(\psoc,\lsoc)$ can be computed analytically by using standard linear algebra. By subtracting the \acrshort{kkt} condition of the coupled \acrshort{qp} subproblem~\eqref{alg::aladin::qp}
\begin{equation}\label{eq::kkt::qp}
    \begin{bmatrix}
        H & A^\top           & J^\top\\
        A &-\frac{I}{\mu}  & 0\\
        J & 0             & 0 
    \end{bmatrix}
    \begin{bmatrix}
        \pqp\\ \lqp -\lambda\\\kqp
    \end{bmatrix}=
    -\begin{bmatrix}
        A^\top\lambda+g\\Ax-b\\0
    \end{bmatrix}
\end{equation}
with identity matrix $I$, and the \acrshort{kkt} condition of the second-order correction subproblem~\eqref{alg::aladin::compensated::qp}
\begin{equation}\label{eq::kkt::soc}
    \begin{bmatrix}
        H & A^\top           & J^\top\\
        A &-\frac{I}{\mu} & 0\\
        J & 0             & 0 
    \end{bmatrix}
    \begin{bmatrix}
        \psoc\\ \lsoc -\lambda\\\ksoc
    \end{bmatrix}=
    -\begin{bmatrix}
        A^\top\lambda+g\\Ax-b\\ r
    \end{bmatrix},
\end{equation}
we obtain a linear system
\begin{equation}\label{eq::soc::equivalent}
    \begin{bmatrix}
        H & A^\top           & J^\top\\
        A &-\frac{I}{\mu} & 0\\
        J & 0             & 0 
    \end{bmatrix}
    \begin{bmatrix}
        \Delta p\\ \Delta \lambda\\ \Delta \kappa
    \end{bmatrix}=
    -\begin{bmatrix}
        0\\0\\r
    \end{bmatrix}
\end{equation}
with difference of these two steps
\begin{equation}
    (\Delta p, \Delta \lambda, \Delta \kappa)=(\psoc, \lsoc, \ksoc)-(\pqp, \lqp, \kqp).
\end{equation}
Under a mild assumption that the \acrshort{kkt} point is regular, we can further reduce the system dimension by eliminating $\Delta \kappa$ 
\begin{equation}
    \begin{bmatrix}
        H & A^\top           \\
        A &-\frac{1}{\mu} \\
    \end{bmatrix}
    \begin{bmatrix}
        \Delta p\\ \Delta \lambda
    \end{bmatrix}=
    -\begin{bmatrix}
         J^\top\cdot \left(J \, M \, J^\top\right)^{-1}\\0
    \end{bmatrix} \cdot r
\end{equation}
with invertible matrix $ M = \left(H+\mu A^\top A \right)^{-1}$. 
\begin{definition}\label{def::regularKKT}
    A \acrshort{kkt} point for a standard constrained optimization problem is regular~\cite{nocedal2006numerical} if \acrfull{licq}, strict complementarity condition and second order sufficient condition are satisfied.
\end{definition}
As a result, the solution to the second-order correction subproblem~\eqref{alg::aladin::compensated::qp} can be computed by
\begin{align}
    \begin{bmatrix}
        \psoc\\ \lsoc
    \end{bmatrix} 
     &=\begin{bmatrix}
        \pqp\\ \lqp
    \end{bmatrix} +
    \begin{bmatrix}
        \Delta p\\ \Delta \lambda
    \end{bmatrix} \notag\\
    &= \begin{bmatrix}
        \pqp\\ \lqp
    \end{bmatrix} -
    \begin{bmatrix}
         I\\ \mu A
    \end{bmatrix}\cdot M \, J^\top \cdot \left(J \, M \, J^\top\right)^{-1} \cdot r.
\end{align}

\begin{remark}
    Locally, if the regularity condition (Definition~\ref{def::regularKKT}) holds, the dual Hessian $JMJ^\top$ is invertible. However, in practice, the  \acrshort{licq} might be violated such that matrix $JMJ^\top$ might not be invertible. In such case, the pseudo-inverse of $JMJ^\top$ has to be used. 
\end{remark}

\subsection{Local Convergence Analysis}
To validate that the local convergence rates are preserved after introducing the correction step, we examine the local convergence property of Algorithm~\ref{alg::aladin} as detailed in the subsequent analysis. 
\begin{theorem}\label{thm::convergence}
Let $(z^*,\lambda^*,\kappa^*)$ be a regular \acrshort{kkt} point for the problem~\eqref{eq::formulation}, let $f$ and $h$ be twice continuously differentiable, and let $\rho \Sigma_\ell$ being sufficiently large for all $\ell\in\mathcal{R}$ so that 
\begin{equation}\label{eq::convexification::Hessian}
    \forall\ell\in\mathcal{R},\;\nabla^2 \left\{ f_\ell(x_\ell) + \kappa_\ell^\top h_\ell(x_\ell)\right\} + \rho \Sigma_\ell\succ 0
\end{equation}
are satisfied. Additionally, let the Hessian approximation $H_\ell$ be accurate enough so that
\begin{equation}\label{eq::Hessian::approximation}
    H_\ell = \nabla^2 \left\{f_\ell(x_\ell) + \kappa_\ell^\top h_\ell(x_\ell)\right\}+\mathcal{O}\left(\norm{x_\ell-z_\ell}\right)
\end{equation}
holds for all $\ell\in\mathcal{R}$. The iterate $(x,\lambda)$ given by Algorithm~\ref{alg::aladin} converges locally to $(z^*,\lambda^*)$ at a quadratic rate.
\end{theorem}
The proof of Theorem~\ref{thm::convergence}  can be established the proof in two steps: we first analyze the convergence rate of Algorithm~\ref{alg::aladin} without the second-order correction step~\ref{alg::aladin::s6}, then we prove that the convergence rate can be preserved if the step is executed during the iterations. The detailed proof can be found in the Appendix.

\begin{remark}
    The term $\mathcal{O}(\norm{x_\ell-z_\ell})$ in~\eqref{eq::Hessian::approximation} is introduced to represent some regularization term used for numerical robustness. Despite these heuristic tricks for regularization, the locally quadratic convergences can be always observed in the sense of verifying the condition~\eqref{eq::Hessian::approximation} numerically.
\end{remark}




\section{Case Study}\label{sec::results}
This section evaluates the performance of the proposed distributed approach (\acrshort{aladincor}) in solving distributed AC \acrshort{opf} for \acrshort{itd} systems~\eqref{eq::formulation}, demonstrating its effective integration of the strengths of its predecessors, i.e., standard \acrshort{aladin} and \acrshort{dcc}, while mitigating their individual limitations. 


\subsection{Simulation Setting}
Three \acrshort{itd} test cases with varying problem sizes and grid topologies are generated based on the standard IEEE test systems. In Case 1, one IEEE 39-bus transmission grid is connected by three IEEE 15-bus radial distribution grids. Case 2 comprises one IEEE 118-bus transmission grid and 20 IEEE 33-bus radial distribution grids. Both cases adopt a star-shaped topology, where the transmission grid is the central hub, and the distribution grids are connected to it. In contrast, Case 3 explores a more complex scenario, where 4 IEEE 118-bus transmission grids are interconnected, resulting in a meshed topology. Additionally, 5 IEEE 33-bus radial distribution grids are connected to each transmission grid. The variation in problem size and grid topology enables a comprehensive analysis and comparison of different approaches under different operational conditions and system complexities.

For a fair comparison, all the algorithms are initialized with a flat start. Following~\cite{Engelmann2019}, the quantities in the following are used to illustrate the convergence behavior
\begin{enumerate}
\item The deviation of optimization variables from the optimal value $\norm{x-x^*}_2$.
\item The primal residual, i.e., the violation of consensus constraint $\norm{A x }_2 = \norm{\sum_{\ell\in\mathcal{R}}A_\ell x_\ell}_2$.\label{quantity::consensus} 
\item The dual residual, i.e. the weighted euclidean distance of the \acrshort{aladin} local step, $\norm{\Sigma(x -z)}_2$. 
\item The solution gap calculated as $\frac{f(x^*)-f(x)}{f(x^*)}$, where $f(x^*)$ is provided by the centralized approach.
\end{enumerate}
When applying the \acrshort{dcc} method from~\cite{LinChenhui2020}, the second and the third quantities are replaced by the mismatch of coupling variables between subproblems and the difference of upper and lower bound of the total cost for the \acrshort{itd} system respectively. Note that the vector $b$ in consensus constraint is neglected since it is a zero-vector in this specific problem, c.f.~\eqref{eq::example::consensus::matrix}. 

The solution accuracy and the solution gap are defined by $\norm{x-x^*}_2$ and $|\frac{f(x)-f(x^*)}{f(x^*)}|$ respectively, while the conic residual is defined by $\norm{\frac{p^2_{ij}+{q}^2_{ij}}{u_i}-l_{ij}}_2$ for all branches $(i,j)\in\mathcal{L}_{\ell\in\mathcal{R}_D}$ in distribution grids, c.f. the conic constraints~\eqref{eq::bfm::soc}.
\begin{figure*}[htbp!]
    \centering
    \includegraphics[width=0.8\textwidth]{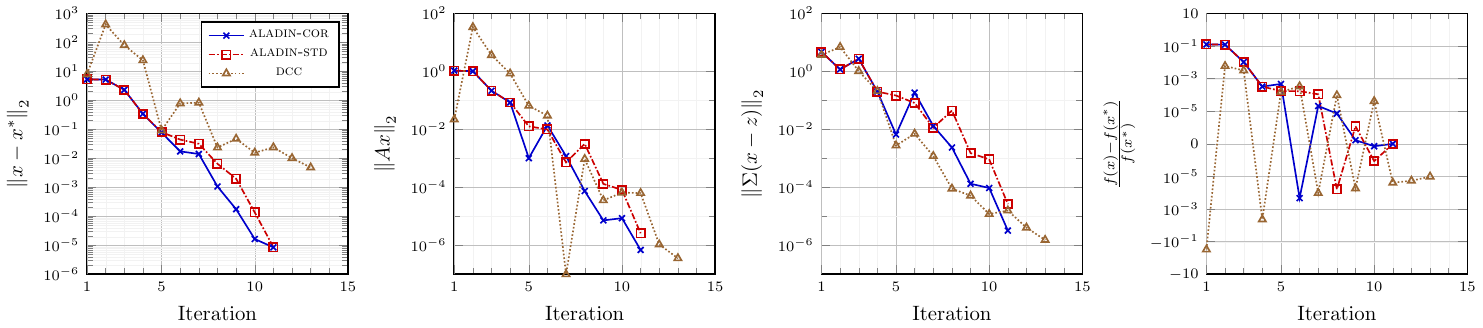}
    \caption{Convergence behavior of different algorithms for Case 1}\label{fig::convergence::case1::alg} \vspace{-4pt}
\end{figure*}
\begin{figure*}[htbp!]
    \centering
    \includegraphics[width=0.8\textwidth]{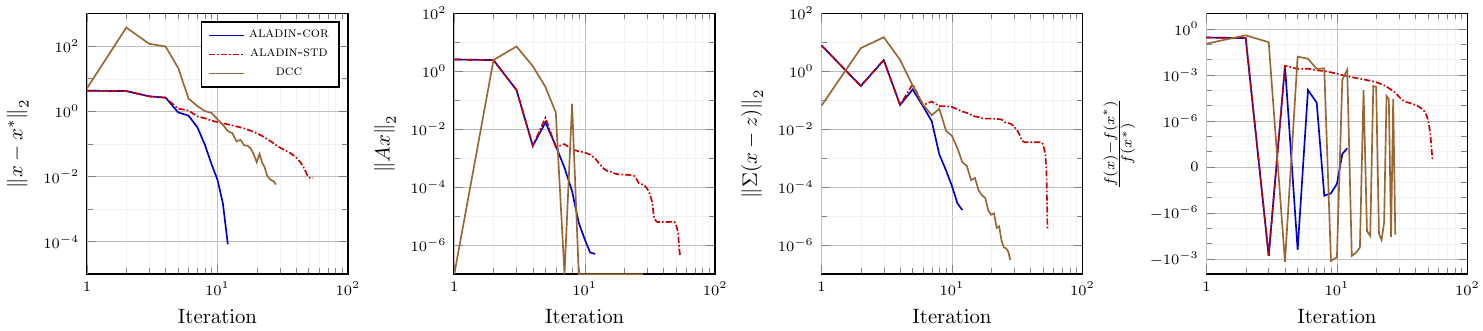}
    \caption{Convergence behavior of different algorithms for Case 2}\label{fig::convergence::case2::alg} \vspace{-4pt}
\end{figure*}
\begin{figure*}[htbp!]
    \centering
    \includegraphics[width=0.8\textwidth]{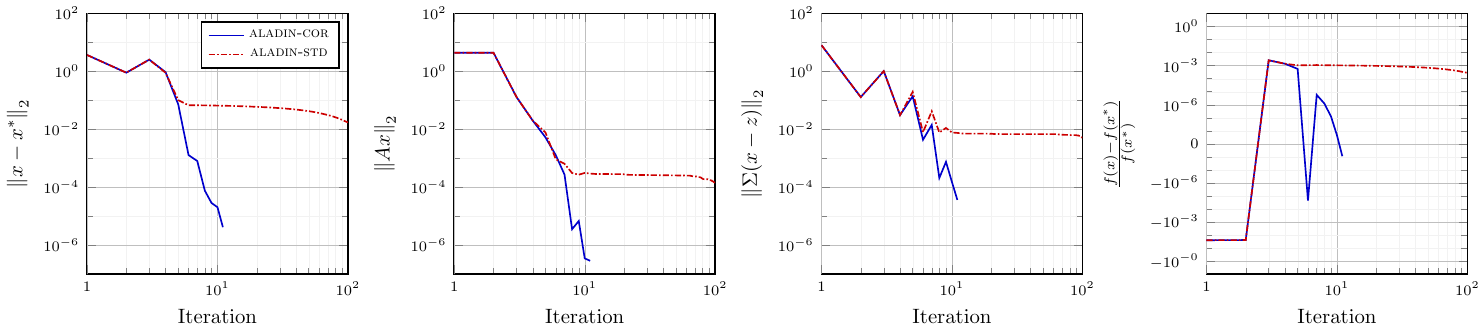}
    \caption{Convergence behavior of different algorithms for Case 3}\label{fig::convergence::case3::alg} \vspace{-4pt}
\end{figure*}
\begin{figure*}[htbp!]
    \centering
    \includegraphics[width=0.8\textwidth]{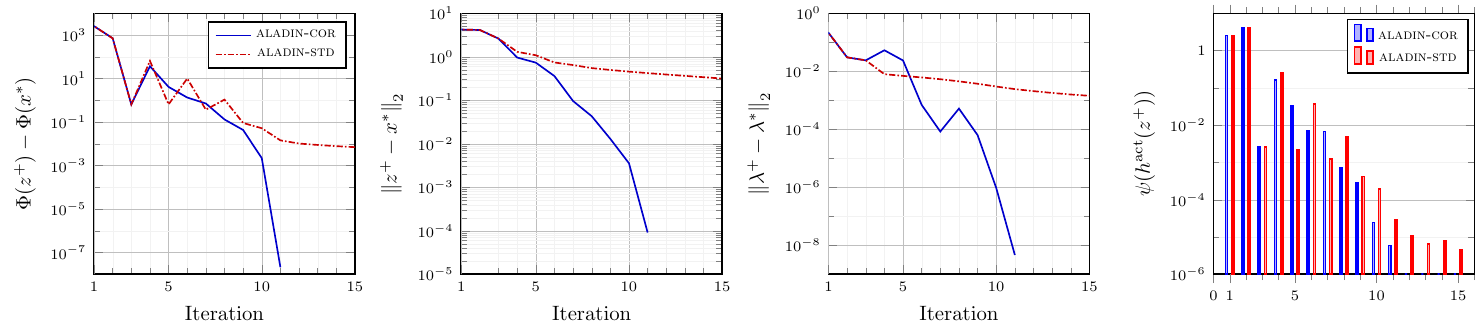}
    \caption{Comparison of the updated primal-dual variables $(z^+,\lambda^+)$ for Case 2}\label{fig::convergence::case2::coordinator}\vspace{-12pt}
\end{figure*}

\subsection{ALADIN-STD vs. ALADIN-COR}\label{sec::aladin::compare}

To demonstrate the improvement of the proposed new \acrshort{aladincor}, we compare its convergence behavior with \acrshort{aladinstd}. In Cases 1 and 2, as illustrated in \autoref{fig::convergence::case1::alg} and \autoref{fig::convergence::case2::alg}, the second-order correction is first carried out at the 4-th iteration of \acrshort{aladincor}. While the impact in the smaller Case 1 is modest, larger Cases 2 and 3, depicted in \autoref{fig::convergence::case2::alg} and \autoref{fig::convergence::case3::alg}, demonstrate significant improvements in convergence rate and solution accuracy using \acrshort{aladincor}. This highlights the scalability and effectiveness of the proposed \acrshort{aladincor} for the nonconvex AC \acrshort{opf} of \acrshort{itd} systems.

For Case 2, a more detailed quality comparison of the primal-dual iterates $(z^+,\lambda^+)$ between \acrshort{aladinstd} and \acrshort{aladincor} is presented in~\autoref{fig::convergence::case2::coordinator}. The comparison includes the gap of the exact penalty function, the deviation of the primal-dual iterates to the local optimizer respectively, and the violation of the active constraint. At the $4$-th iteration, the correction step is first activated due to the increase in the merit function~\eqref{eq::merit} and significant violation of the constraints~\eqref{ineq::nonlinear} as shown in \autoref{fig::convergence::case2::coordinator}. This results in considerable improvement in the objective values and constraint violation, as displayed in \autoref{tb::correction::violation}, contrasting with the damping observed in subsequent iterations of \acrshort{aladinstd}. A comprehensive comparison of algorithm performance is provided in \autoref{TB::numerical-result}.


The second-order correction strategy presents a minimal trade-off in its implementation. 
The activation of the correction step is reserved where there is an increase in the merit function accompanied by significant constraint violations, as evidenced in only a few iterations shown in \autoref{tb::tradeoff} for all three cases. Additionally, the computational and communication demands for executing each second-order correction step are substantially mitigated through the reuse of matrix inverses and factorizations, as discussed in Section~\ref{sec::alg::practice}. Requiring only basic linear algebra for computation, the additional computing time $t^\textsc{soc}$ is negligible, comprising around $5\%$ total computing time (\autoref{tb::tradeoff}).



\begin{table}[htbp!]
    \centering
    \scriptsize
    \caption{Comparison of the $4$-th iterates for Case~2}\label{tb::global::step}
    \begin{tabular}{cccc}
    \toprule
        & $\frac{f(z)-f(x^*)}{f(x^*)}$ & $\norm{A z}_1$ & $\psi(h^\textrm{act}(z))$ \\\midrule
       $\zqp$ & $2.9\times10^{-3}$ & $1.5\times10^{-17}$ & $0.257$\\
       $\zsoc$ & $-1.3\times10^{-5}$ & $2.0\times10^{-17}$ & $0.163$ \\\bottomrule
    \end{tabular} 
\label{tb::correction::violation}
\end{table}
\begin{table}[htbp!]
    \centering
    \scriptsize
    \caption{Computing time by \acrshort{aladincor}}\label{tb::tradeoff}
    \begin{tabular}{cccccc}
    \toprule
        Case & Iterations & Corrected Iterations &  $t^\textsc{total} [s]$& $t^\textsc{soc}[s]$ \\\midrule
        1 &11&2& $0.507$  & $ 4.36\times 10^{-3}$\\
        2 &12&4& $3.222$ & $7.61 \times 10^{-2}$ \\
        3 &11&4& $5.044$ & $8.98\times 10^{-2}$\\\bottomrule
    \end{tabular}
\label{tb::correction::time}
\end{table}
This correction strategy, balancing minimal additional computational requirements with significant advancements in convergence, significantly improves the practicality and effectiveness of the \acrshort{aladin}-type algorithm. Consequently, \acrshort{aladincor} stands out as a robust and efficient solution for distributed AC \acrshort{opf} of \acrshort{itd} systems, delivering high-performance outcomes with minimal communication and computational overheads.

\begin{table*}[ht]
    \caption{Comparisons of different algorithms} \label{TB::numerical-result}
    \centering
    \footnotesize
    \renewcommand{\arraystretch}{1.4}
    \begin{tabular}{ccccccccc}\toprule
        Case & Number of Buses & Number of Regions &  Algorithm &  Iterations & Time [s]   &$\norm{x-x^*}_2$& Solution Gap & Coic Residual\\\toprule
        \multirow{3}{*}{1} & \multirow{3}{*}{84} & \multirow{3}{*}{4} & \acrshort{dcc} &  $13$ & $2.881$ & $4.84\times10^{-3}$ & $1.02\times10^{-5}$ & $ 1.98\times10^{-7}$\\
        & &&\acrshort{aladinstd} & $11$ & $0.503$ &  $8.62\times 10^{-6}$ & $4.91\times 10^{-8}$ & $1.08\times 10^{-5}$\\
        & &&\acrshort{aladincor}  &$11$ & $0.507$ &  $8.52\times 10^{-6}$ & $1.91\times 10^{-8}$ & $1.08\times 10^{-5}$\\\hline
        \multirow{3}{*}{2} &\multirow{3}{*}{778} & \multirow{3}{*}{21} &\acrshort{dcc} & $28$ & $22.545$ & $5.65\times10^{-3}$ & $2.56\times 10^{-5}$ & $2.68\times10^{-7}$ \\
        && &\acrshort{aladinstd}  & $55$ & $14.909$ & $8.95\times 10^{-3} $ & $2.84\times 10^{-9}$ & $4.95\times 10^{-3}$\\
        && &\acrshort{aladincor}  & $12$ & $3.222$ &  $8.36\times 10^{-5}$ & $1.61\times 10^{-8}$ & $3.02\times 10^{-5}$\\\hline
        \multirow{3}{*}{3} &  \multirow{3}{*}{1132} & \multirow{3}{*}{24} & \acrshort{dcc} &  \multicolumn{5}{c}{Did not converge}\\
        &&&\acrshort{aladinstd} & $100$& $43.569$& $1.73\times 10^{-2}$ & $2.78\times 10^{-4}$& $1.47\times 10^{-3}$\\
        && &\acrshort{aladincor} & $11$ & $ 5.044$ & $4.23\times 10^{-6}$ &  $8.81\times 10^{-9}$& $8.32\times10^{-6}$ \\\bottomrule
    \end{tabular}\vspace{-15pt}
\end{table*}

\subsection{DCC vs. ALADIN-COR}

The recently proposed \acrshort{dcc} framework~\cite{LinChenhui2020} employs a two-layer strategy for solving AC \acrshort{opf} problems of \acrshort{itd} system~\eqref{eq::formulation}. It solves transmission and distribution subproblems sequentially---distribution subproblems are solved in parallel in the lower layer by local agents, using conic optimization with an $L_1$ penalty for coupling mismatches. Additionally, the local agents generate quadratic approximations and lower bounds for distribution costs through tangent planes with respect to coupling variables. The upper layer formulates a nonconvex AC \acrshort{opf} problem for transmission, integrating distribution costs approximated from the lower layer. The transmission grid serves as a centralized coordinator, optimizing and sharing coupling variables with local distributions, thus preserving privacy since detailed data isn't exchanged.

\begin{figure}[htbp!]
    \centering
    \includegraphics[width=0.48\textwidth]{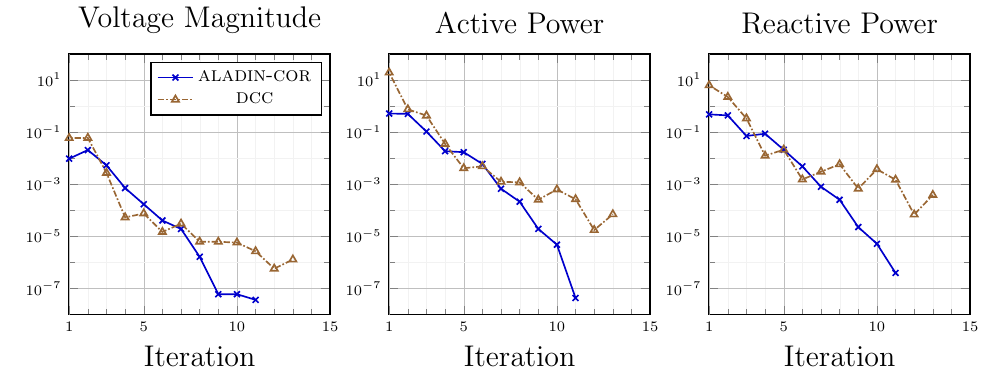}
    \caption{\acrshort{rmse} of coupling variables for Case 1}\label{fig::convergence::case1::vpq} 
\end{figure}

\begin{figure}[htbp!]
    \centering
    \includegraphics[width=0.48\textwidth]{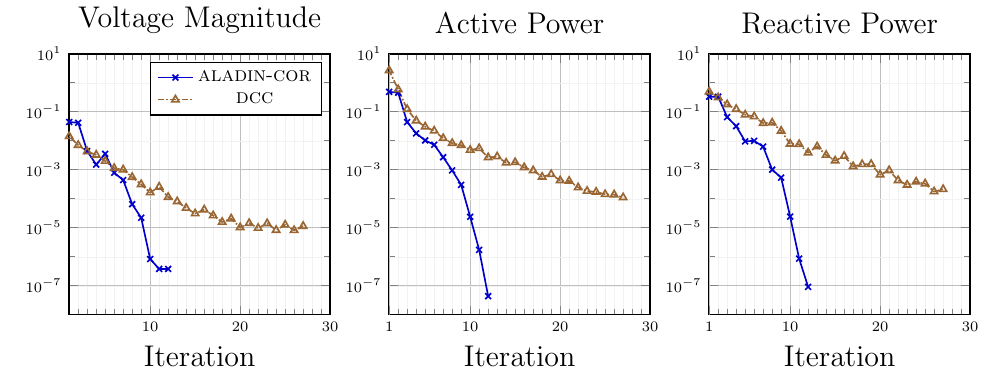}
    \caption{\acrshort{rmse} of coupling variables for Case 2}\label{fig::convergence::case2::vpq} 
\end{figure}

Although the two-layer strategy provides convergence guarantees in~\cite{LinChenhui2020}, some limitations need to be acknowledged. It is primarily designed for star-shaped \acrshort{itd} systems with multiple radial distributions connecting a single transmission, restricting its adaptability to more diverse and realistic topologies involving, for instance, multiple transmission grids, meshed distribution grids, or multiple transmission grids interconnected in a meshed topology. Moreover, its effectiveness depends on the solvability of one nonconvex transmission subproblem in the upper layer, posing issues regarding numerical robustness and scalability. Furthermore, \acrshort{dcc} can lead to numerical instability or degeneracy in cases of weakly active conic constraints. Such situations often result in zigzagging in coupling variables when near high-precision solutions, as observed in~\autoref{fig::convergence::case1::vpq} and~\autoref{fig::convergence::case2::vpq}.

In contrast, our proposed \acrshort{aladincor} is tailored for generic distributed optimization, treating all subproblems equitably and providing convergence guarantees for \acrshort{itd} systems with more diverse and realistic topology. It incorporates the proximal regularization method and the conic relaxation method discussed to improve the computation efficiency of local subproblems while strategically implementing second-order corrections to enhance numerical stability while preserving quadratic convergence rates.

To validate our approach's effectiveness, we rigorously compare with the study in~\cite{LinChenhui2020}, focusing on the \acrfull{rmse} of coupling variables, as illustrated for Case 1 in~\autoref{fig::convergence::case1::vpq} and Case 2 in~\autoref{fig::convergence::case2::vpq}. These demonstrate that our case studies achieved an accuracy level comparable to, or even surpassing, those in the original study. 
Notably, the \acrshort{dcc} approach exhibits significant decision variable deviations in early iterations, as shown in ~\autoref{fig::convergence::case1::alg},~\autoref{fig::convergence::case2::alg}, and minor oscillations near local minimizers in terms of \acrshort{rmse} of coupling variables, as shown in~\autoref{fig::convergence::case1::vpq},~\autoref{fig::convergence::case2::vpq}.
In contrast, our proposed \acrshort{aladincor} consistently demonstrates rapid convergence and also outperforms \acrshort{dcc} in managing the complex Case 3, which involved four interconnected transmission grids in a mesh topology.

\vspace{-2pt}
\vspace{-2pt}
 
These results underscore the superior scalability and effectiveness of the proposed \acrshort{aladincor} in tackling the challenges of AC \acrshort{opf} in various \acrshort{itd} systems. Our approach achieves not only higher accuracy than the original study of the \acrshort{dcc} approach but also demonstrates enhanced performance and efficiency, validating its robustness for diverse system topologies.

\section{Conclusions \& Outlook}\label{sec::conclusions}

The present paper proposes a novel \acrshort{aladincor} algorithm for efficiently solving the distributed AC \acrshort{opf} problem of \acrshort{itd} systems regarding complex topology. Our detailed analysis and proof confirm that the proposed \acrshort{aladin}-type algorithm retains a locally quadratic convergence rate even with the added second-order correction step, ensuring its efficiency and numerical robustness. Numerical experiments conducted on three \acrshort{itd} benchmark cases varying in problem sizes and grid topologies, showcased its effective integration of the strengths of its predecessors, i.e., standard \acrshort{aladin} and \acrshort{dcc}, while mitigating their individual limitations. Consequently, the proposed \acrshort{aladincor} approach offers a more stable, efficient, and scalable solution compared with its predecessors.

The proposed methodology opens several problems and extensions for future research. These include exploring the impacts of communication delays, packet losses, and asynchronous updates on the algorithm's performance, ideally within a distributed computing software architecture, ideally within a distributed software architecture or parallel computing toolbox. Such studies aim to fortify the algorithm's resilience in actual power systems, where communication constraints are present. Further directions include scaling the algorithm for large-scale power systems, integrating a distributionally robust framework to accommodate uncertainties in renewable energy, and employing a receding horizon approach for online scheduling.


\appendix
\label{appendix}
\footnotesize
According to the assumptions of regularity and $\rho$, the local minimizer of subproblems~\eqref{alg::aladin::nlp}, $x_\ell$ is parametric with $(z,\lambda)$ and the solution maps are Lipschitz continuous, i.e.,
there exists constants $\chi_1,\chi_2>0$ such that 
\begin{equation}\label{eq::Lipschitz}
\left\|x-z^*\right\|\leq  \chi_1\left\|z-z^*\right\| + \chi_2\left\|\lambda-\lambda^*\right\|.
\end{equation}
From the local convergence analysis of Newton methods~\cite[Ch.~3.3]{nocedal2006numerical}, we have
\begin{align}
\norm{\begin{bmatrix}
\zqp-z^*\\
\lqp-\lambda^*\end{bmatrix}}\leq &\norm{H-\nabla^2 \left\{f(x) + \kappa^\top h(x)\right\}}\cdot\mathcal{O}\left(\norm{x-z^*}\right)\notag\\ &+\mathcal{O}(\norm{x-z^*}^2).	\notag
\end{align}
By considering the accuracy of the Hessian approximation~\eqref{eq::Hessian::approximation}, the quadratic contractions
\begin{subequations}\label{eq::contractions}
    \begin{align}
        \norm{\zqp-z^*}\leq \mathcal{O}&(\norm{x-z^*}^2),\\
        \norm{\lqp-\lambda^*}\leq \mathcal{O}&(\norm{x-z^*}^2),
    \end{align}
\end{subequations}
can be established. By combining~\eqref{eq::Lipschitz} and~\eqref{eq::contractions}, locally quadratic convergence of  Algorithm~\ref{alg::aladin} can be guaranteed if the second-order correction step~\ref{alg::aladin::s5} is never executed, i.e., $(\zqp,\lqp)$ is always accepted.

Then, we take the second-order correction step into consideration. Following~\cite{fletcher1982second}, utilizing the relations $\zqp-x=\pqp$ yields an upper bound of the step $\pqp$ 
\begin{align}
\norm{\pqp} & = \norm{\zqp - x} \notag   \\ 
  & \leq  \norm{\zqp - z^*} + \norm{x-z^*} \notag   \\ 
  & \leq  \mathcal{O}(\norm{x-z^*}^2) + \mathcal{O}(\norm{x-z^*}).\label{eq::upperbound::p}
\end{align}
By Taylor series, we have
\begin{align}\label{eq::soc::constraint::residual}
    r_\ell  &= h^\textrm{act}_{\ell}(x_{\ell}+\pqp_\ell) =  h^\textrm{act}_{\ell}(x_{\ell}) + J_\ell \pqp_\ell + \mathcal{O}(\norm{\pqp_\ell}^2)\notag\\
    & = \mathcal{O}(\norm{\pqp_\ell}^2),\;\forall \ell\in\mathcal{R},
\end{align}
where iterates $x_\ell$ satisfies $h^\textrm{act}_\ell(x_\ell)=0$ and the step $\pqp_\ell$ satisfies the linearized equality constraint~\eqref{eq::active}.
Under the regularity condition (Definition~\ref{def::regularKKT}), the \acrshort{kkt} matrix in the right-hand side of~\eqref{eq::soc::equivalent} is invertible, and the corresponding inverse matrix is bounded. Consequently, by combining~\eqref{eq::soc::equivalent},~\eqref{eq::upperbound::p}, \eqref{eq::soc::constraint::residual}, we have a quadratic contraction of primal and dual variables
\begin{equation} \label{eq::contraction::soc}
    \norm{\begin{bmatrix}
         \psoc - \pqp\\
         \lsoc - \lqp\\
    \end{bmatrix}} \leq  \mathcal{O}(\norm{r}) \leq \mathcal{O}(\norm{x-z^*}^2).
\end{equation}

From~\eqref{eq::Lipschitz}, \eqref{eq::contractions} and \eqref{eq::contraction::soc}, it is sufficient to prove that the locally quadratic convergence can be achieved no matter whether the second-order correction step (step~\ref{alg::aladin::s5}) is executed during the iterations.
\bibliographystyle{ieeetr}
\bibliography{new_bib}



\end{document}